\input harvmac

\def \lr {\lref }

\def \s {\sigma}
\def \r {\rho}

\def \const {{\rm const }}

\def\F{{\cal F }}
\def \ov {\over}


\rightline{ECM-UB-03/14}

\baselineskip8pt
\Title{\vbox {\baselineskip 6pt{\hbox{ }}{\hbox {
}}{\hbox {}}{\hbox{ }} {\hbox{ }}} } {\vbox{\centerline {
Cosmological string models from Milne spaces} \vskip4pt
\centerline{and $SL(2,Z)$ orbifold} }} \vskip -20 true pt
\centerline { {Jorge G. Russo \footnote {$^*$} {e-mail address:
jrusso@mail.cern.ch} }}

\medskip \smallskip

\centerline {\it {} Departament ECM, Facultat de F\'\i sica,}
\smallskip
\centerline {\it Universitat de Barcelona, Diagonal 647, E-08028,
Barcelona, Spain}
\smallskip
\centerline {\it Instituci\' o Catalana de Recerca i Estudis
Avan\c cats (ICREA)}

\bigskip
\centerline {\bf Abstract}
\medskip
\baselineskip10pt \noindent

The $n+1$-dimensional Milne Universe with extra free
directions is used to construct simple FRW cosmological string
models in four dimensions, describing expansion in the presence of
 matter with $p=\kappa \rho $, $\kappa=(4-n)/3n$. We then consider
 the $n=2$ case and make  $SL(2,{\bf Z})$ orbifold identifications.
The model is surprisingly related to the null orbifold with an extra
reflection generator.
The study of the string spectrum
involves the theory of harmonic functions in the fundamental domain of $SL(2,{\bf Z})$.
In particular, from this theory one can deduce a bound for the energy gap
and the fact that there are an infinite number of excitations
with a finite degeneracy.
 We discuss the structure of wave functions and give examples of
physical winding states becoming light
near the singularity.

\medskip

\Date {May 2003}
\noblackbox
\baselineskip 16pt plus 2pt minus 2pt


\noblackbox \overfullrule=0pt



\lref\HS{G.~T.~Horowitz and A.~R.~Steif, ``Singular
String Solutions With Nonsingular Initial Data,'' Phys.\ Lett.\ B
{\bf 258}, 91 (1991).
}

\lref\russoc{J.~G.~Russo,
``Construction of SL(2,{\bf Z}) invariant amplitudes in type IIB superstring  theory,''
Nucl.\ Phys.\ B {\bf 535}, 116 (1998)
hep-th/9802090.}

\lref\LMS{
H.~Liu, G.~Moore and N.~Seiberg,
``Strings in a time-dependent orbifold",
JHEP 0206 (2002) 045, hep-th/0204168;
`` Strings in time-dependent orbifolds",
JHEP 0210 (2002) 031, hep-th/0206182.
}

\lr\CCK{L.~Cornalba, M.~S.~Costa and C.~Kounnas,
``A resolution of the cosmological singularity with orientifolds",
Nucl.\ Phys.\ B {\bf 637} (2002) 378,
hep-th/0204261.}

\lr\CKR{
B.~Craps, D.~Kutasov and G.~Rajesh,
``String propagation in the presence of cosmological singularities",
JHEP 0206 (2002) 053,
hep-th/0205101.}

\lr\turok{C.~Gordon and N.~Turok,
``Cosmological perturbations through a general relativistic bounce",
hep-th/0206138.}

\lr\Fabi{
M.~Fabinger and J.~McGreevy,
``On smooth time-dependent orbifolds and null singularities",
hep-th/0206196.}

\lr\HP{
G.~T.~Horowitz and J.~Polchinski,
``Instability of spacelike and null orbifold singularities",
Phys.\ Rev.\ D {\bf 66} (2002) 103512,
hep-th/0206228.}

\lr\EGKR{S.~Elitzur, A.~Giveon, D.~Kutasov and E.~Rabinovici,
``From big bang to big crunch and beyond",
JHEP 0206 (2002) 017,
hep-th/0204189.}

\lref\TW{P.~K.~Townsend and M.~N.~Wohlfarth,
``Accelerating cosmologies from compactification,''
hep-th/0303097.}

\lref\KR{A.~Kehagias and J.~G.~Russo,
``Hyperbolic spaces in string and M-theory,''
JHEP {\bf 0007}, 027 (2000),
hep-th/0003281.}

\lref\FY{

Y.~Fujii and K.~Yamagishi,
``Killing Spinors On Spheres And Hyperbolic Manifolds,''
J.\ Math.\ Phys.\  {\bf 27}, 979 (1986).}

\lref\LPR{H.~Lu, C.~N.~Pope and J.~Rahmfeld,
``A construction of Killing spinors on S**n,''
J.\ Math.\ Phys.\  {\bf 40}, 4518 (1999),
hep-th/9805151.}

\lref\terras{A. Terras, ``Harmonic analysis on symmetric spaces and applications," Vol. I, Springer-Verlag, 1985.}

\lref\yamamoto{
K.~Yamamoto, T.~Tanaka and M.~Sasaki,
``Particle spectrum created through bubble nucleation and quantum field theory in the Milne Universe,''
Phys.\ Rev.\ D {\bf 51}, 2968 (1995)
[arXiv:gr-qc/9412011];
T.~Tanaka and M.~Sasaki,
``Quantized gravitational waves in the Milne universe,''
Phys.\ Rev.\ D {\bf 55}, 6061 (1997)
[arXiv:gr-qc/9610060].}

\lref\Simon{ J.~Simon, ``The geometry of null rotation
identifications,'' JHEP {\bf 0206}, 001 (2002),
hep-th/0203201.
}

\lr\CC{  L.~Cornalba and M.~S.~Costa,
``A New Cosmological Scenario in String Theory",
Phys.\ Rev.\ D {\bf 66} (2002) 066001,
hep-th/0203031.}

\lref\tolley{ A.~J.~Tolley and N.~Turok, ``Quantum fields in a
big crunch / big bang spacetime,'' arXiv:hep-th/0204091.
}

\lref\lawren{A.~Lawrence, ``On the instability of 3d null
singularities'', JHEP 0211 (2002) 019, hep-th/0205288.}

\lref\bala{ V.~Balasubramanian, S.~F.~Hassan,
E.~Keski-Vakkuri and A.~Naqvi, ``A space-time orbifold: A toy
model for a cosmological singularity,''
Phys.\ Rev.\ D {\bf 67}, 026003 (2003), hep-th/0202187.
}

\lref\nekra{ N.~A.~Nekrasov, ``Milne universe, tachyons, and
quantum group,'' hep-th/0203112.
}

\lr\BCKR{
M.~Berkooz, B.~Craps, D.~Kutasov and G.~Rajesh,
``Comments on cosmological singularities in string theory", JHEP {\bf 0303}, 031 (2003),
hep-th/0212215.}

\lr\FH{
M.~Fabinger and S.~Hellerman,
``Stringy resolutions of null singularities",
hep-th/0212223.}

\lr\EGR{
S.~Elitzur, A.~Giveon and E.~Rabinovici,
``Removing singularities",
JHEP 0301 (2003) 017,
hep-th/0212242.}

\lr\silva{S.~Nasri, P.~J.~Silva, G.~D.~Starkman and M.~Trodden,
``Radion stabilization in compact hyperbolic extra dimensions,''
Phys.\ Rev.\ D {\bf 66}, 045029 (2002),
hep-th/0201063.}

\lref \rrt { J.~G.~Russo and A.~A.~Tseytlin, ``Magnetic flux tube
models in superstring theory,'' Nucl.\ Phys.\ B {\bf 461}, 131
(1996) [hep-th/9508068].
}


\lr \schwarz{ J.~H.~Schwarz, ``Covariant Field Equations Of Chiral
N=2 D = 10 Supergravity,'' Nucl.\ Phys.\ B {\bf 226}, 269 (1983).
}

\lr\KOSST{J.~Khoury, B.~A.~Ovrut, N.~Seiberg,
P.~J.~Steinhardt and N.~Turok, ``From big crunch to big bang,''
Phys.\ Rev.\ D {\bf 65}, 086007 (2002),
hep-th/0108187;
N.~Seiberg, ``From big crunch to big bang - is it possible?,''
hep-th/0201039.
}

\lr\PRT{G.~Papadopoulos, J.~G.~Russo and A.~A.~Tseytlin,
``Solvable model of strings in a time-dependent plane-wave background,''
Class.\ Quant.\ Grav.\  {\bf 20}, 969 (2003),
hep-th/0211289.}

\lr\costa{L.~Cornalba and M.~S.~Costa,
``On the classical stability of orientifold cosmologies,''
hep-th/0302137.}



\newsec{Introduction}

Deep questions in cosmology, such as the nature of the big bang
singularity or initial boundary conditions, need to be addressed
in the context of a quantum gravity theory.
 Understanding string theory in time-dependent
backgrounds leads to a number of technical problems which have
been considered in many recent works \refs{\KOSST\bala\CC \nekra\Simon\LMS\EGKR\CCK\CKR\lawren\Fabi\HP\PRT\BCKR\FH - \costa }.
 In particular, in \LMS\ the null  orbifold model \HS  --~a
time dependent geometry consisting in flat space with some
identifications~-- was studied in detail. The string model
of this Lorentzian orbifold
contains some instabilities which are reflected in IR divergences
in scattering amplitudes. Their physical origin is a large back reaction in
the geometry as soon as particles are introduced \refs{\LMS, \lawren , \HP }.

In the present work we will be interested in the general class of
time-dependent locally flat spacetimes obtained from the $n+1$ dimensional Milne
Universe. They  can be viewed as a particular case  of  hyperbolic compactification in
string theory (see \KR ), which have recently attracted some interest as
they lead to interesting cosmologies \TW .
We will consider in more detail the  $n=2$ case, which, with proper identifications, is related to the null orbifold. Here we make orbifold identifications on the upper half plane by using the (elliptic) modular group $SL(2,{\bf Z})$.

Milne spaces in the context of inflationary cosmology were studied in
\yamamoto . String models in 1+1 dimensional Milne space were discussed
in \refs{\HS, \KOSST , \CC, \nekra }. Discussions on higher dimensional Milne spaces can be found in \refs{\HS , \KR }, and
more recently  in \costa  .

In section 2 we discuss the simplest string cosmology based on
flat Milne space. We will see that already these simple backgrounds can be used to obtain  interesting cosmologies in four dimensions.
This section makes a connection with recent work on
hyperbolic compactification \TW , and it can be viewed as a physical motivation for string models based on Milne space.
In section 3 we introduce the $SL(2,{\bf Z})$ orbifold model.
The study of this model
involves harmonic analysis on symmetric spaces and number theory,
from which we will extract some results for the string spectrum.
 We also discuss the construction of wave functions and physical string states.

\newsec{String cosmology from $n+1$-dimensional Milne Universe}

The $n+1$-dimensional Milne space is described by the metric
\eqn\miln{ds^2=-dt^2+t^2 dH_n^2\ ,} where $dH_n^2$ is the arc
element of the hyperboloid or upper half $n$ plane. In Poincar\'e
coordinates, it is \eqn\hyp{dH_n^2= {1\over
z^2}(dz^2+dz_1^2+...+dz_{n-1}^2)\ . } The space is flat, as it is
evident upon introducing Cartesian coordinates as follows:
\eqn\cartes{U= {t\ov z} \ ,\ \ \ \  V=tz+{t\ov z}\sum_{i=1}^{n-1}
z_i^2 \ ,\ \ \ \ X_i={tz_i\ov z} \ .} This provides the embedding
of the hyperboloid in $n+1$ Minkowski space,
$$
ds^2=-dUdV+dX_i^2 \ ,
$$
where the hyperboloid is described by \eqn\emb{ t^2 =
UV-X_1^2-...-X_{n-1}^2 \ ,} which exhibits the $SO(1,n)$ isometry
of $H_n$.

Alternatively, one can use angular or Milne coordinates
\eqn\hypp{
dH_n^2= d\rho^2+\sinh^2\rho\ d\Omega_{n-1}^2 \ ,
}
and
define \eqn\eee{ T=t \cosh\rho \ ,\ \ \ \ R=t\sinh\rho \ ,} in
terms of which the metric \miln\ reads \eqn\mmn{
ds^2=-dT^2+dR^2+R^2 d\Omega_{n-1}^2\ . }

 Now let us consider
type II string theory in the following background
\eqn\aaa{ds^2_{10} = -dt^2+t^2
dH_n^2+dx_1^2+dx_2^2+dx_3^2+dy_1^2+...+dy_{6-n}^2 \ , } where the
$y$ coordinates describe compact internal dimensions. All other
supergravity fields are trivial. This is an exact solution (to all
alpha prime orders) of string theory, since the Riemann tensor
identically vanishes (though it may receive quantum string-loop corrections).
Similarly, one can write down the analogous
eleven dimensional metric, which is a solution of M theory.
The internal space described by the $y$ coordinates can be
replaced by any Ricci flat space, giving a more general class of
cosmological backgrounds which are solutions to leading order in $\alpha '$.

 It is surprising that an interesting four dimensional FRW cosmology can be
obtained from the model \aaa . First,  we replace the hyperboloid
$H_n$ by a finite volume space $H_n/\Gamma $, where $\Gamma $ is a
discrete subgroup of $SO(1,n)$ such that the space has finite
volume, as in \refs{\KR,\TW}. Then we compactify to four
dimensions. To obtain the four dimensional Einstein frame metric,
we write the ten-dimensional metric in the form
 \eqn\eins{ds^2=e^{2a(t)}ds^2_{4E}+e^{2b(t)}dH_n^2+dy_1^2+...+dy_{6-n}^2 \ ,
 }
$$
ds^2_{4E}=e^{2c(t)}(-dt^2+dx_1^2+dx_2^2+dx_3^2) \ , $$
 and the
condition that $ds^2_{4E}$ is in the Einstein frame is $e^{2a}e^{n
b}=1$. Comparing to \aaa , one obtains
\eqn\eein{
ds^2_{4E} =
t^n(-dt^2+dx_1^2+dx_2^2+dx_3^2)\ , } or
 \eqn\nnn{ds^2_{4E} =
-d \tau ^2+ \tau ^{2n\over n+2}(dx_1^2+dx_2^2+dx_3^2) \ .
 }
\noindent  This
corresponds to 4d Einstein equations coupled to an energy momentum
tensor of a perfect fluid with \eqn\pff{p=\kappa \r \ ,\ \ \ \
\kappa={4-n\over 3n} \ . } Although we have started with vacuum
Einstein equations in ten dimensions, the four dimensional
Einstein metric describes a homogeneous and isotropic space in
presence of matter. This matter is of course  the scalar field
associated with the modulus representing the volume of the
hyperbolic space. Interestingly, the above metric is the
asymptotic (large time) form of the models of \TW . For $n=4$, it
describes a universe filled with dust, and for $n>4$ a universe
filled with negative pressure matter.

Since the models are based on a flat ten dimensional geometry, the
$n+1$ dimensional Milne Universes provide a simple setup for the
study of interesting cosmological string models.

\medskip

So let us consider strings propagating in this space. An important
question is whether the string model is exactly solvable.
 To start
with, consider the model \aaa\ based on $H_n$ with no
identification, i.e. $\Gamma=1$. The string equations of motion in
Cartesian coordinates \cartes\ are solved explicitly in terms of
right and left moving modes. However,  from the relation \emb \ it
follows that \eqn\condi{ UV-X_i^2\geq 0 \ .}
If the physical space is restricted only to this Milne patch, say with $t>0$, then the string
coordinates are subject to the constraint that the string lives
in the interior of  the future directed  light cone; the space is not geodesically complete and a full description requires boundary conditions at the light cone surface. In
string theory, it is possible that consistency also requires the
inclusion of the past light cone. In this case, the geometry would
describe a universe contracting to a big crunch which makes a
transition to an expanding big bang universe.

 It is
non-trivial to impose the condition \condi\ in string theory.
On the other hand, if the full space $U,V,X_i$
is considered, closed timelike curves can arise in the exterior of light cone as a result of identifications.

We briefly comment on a possible approach with a boundary on the light cone surface.
 A string which at a given instant fully lies
in the interior of the future light cone, will satisfy the
condition $UV-X_i^2>0$ at all future times. Causal propagation implies
that all closed strings in the physical space emerge from the
light cone surface, so there is some interval of time where the
condition $UV-X_i^2>0$ is satisfied only for a piece of the
string (an example is given at the end of section 3). It is natural to describe the strings intersecting the
light cone in terms of open strings with ends attached to the null
light-cone surface, which can be thought of as a ``null D brane".
 In this picture, the set of open
string states attached to the null surface determine the initial
state, i.e. they would be described by some ``in" state in the
open string Hilbert space.
The closed strings living in the
interior of the light cone would originate by emission from the
null D brane. Thus the time evolution of the cosmology would be
determined by the initial state on the null D brane and the decay
properties of this initial state.

\smallskip

 Let us now comment on the supersymmetry of these models.
 The hyperbolic upper half $n$-plane
$H_n$ has Killing spinors transforming in the spinorial
representation of $SO(1,n-1)$. They were explicitly constructed in
\FY \ (see also discussions in \refs{\LPR, \KR, \silva }). The simplest model \aaa\
with $\Gamma=1$ is supersymmetric. When $\Gamma $ is not trivial
and the space $H_n/\Gamma $ has finite volume, it was shown in \KR
\ that for $n=$even supersymmetries are always broken by the
identifications.  For $n=$odd and  $n\geq 5$,  the analysis of
\KR\ does not exclude that an appropriate choice of $\Gamma $
could give a supersymmetric model with finite volume hyperbolic
space. It would be interesting to find some example.

\newsec{$SL(2,{\bf Z})$ orbifold}

Let us consider the case $n=2$ and recall its relation to the null orbifold  \HS . The
metric is
\eqn\metrica{
 ds^2=-d t^2+{ t^2\ov
y ^2}(d x ^2+dy ^2) \ . } The constant time slices describe the
hyperbolic upper half plane $H_2$. Introducing $U= t/y $, $V= ty
$, we obtain \eqn\abc{ ds^2=-dUdV+U^2d x ^2  \ . } This  form
exhibits an orbifold singularity at $U=0$ moving at the speed of
light. Introducing new coordinates \eqn\nuev{ u=U\ ,\ \ \ \ v=V+U
x ^2, \ \ \ \ X=U x  \ ,} we get \eqn\abb{ ds^2=-dudv+dX^2 \ ,\ \
\ \ \ u=X_0-X_1\ ,\ \   v=X_0+X_1\ . } The null (or ``parabolic")
orbifold geometry of \HS , studied by Liu, Moore and Seiberg \LMS
, is obtained by orbifolding the isometry $ x = x +1$. In the
coordinates $( u,v,X )$ this leads to the identification \foot{
In the notation of \LMS\ one has $x^+=u,\ x^-=v/2 $.} \eqn\idenf{
(u,v,X) \equiv (u,v+un^2+2 X n, X+n\ u) }

Now we consider another geometry: we replace $H_2$ in \metrica\ by
$\F =SL(2,{\bf Z})\setminus H_2$.\foot{A brief discussion of this geometry is in section 5  of \russoc .}
$SL(2,{\bf Z})$ transformations are generated by $z\to z+1$
and $z\to -1/z$. The first transformation is precisely what leads
to the null orbifold \idenf .
Now we have the  extra orbifold
identification $ z \to -1/ z $, with $ z = x +iy $
(and $ t\to  t$). This isometry is
surprisingly simple in the coordinates $(u,v,X)$. It just
corresponds to the spatial reflection:
 \eqn\pari{ u\to v\ ,\ \
\ \ v\to u\ ,\ \ \ \ X\to -X \ ,
}
where we have used the relation
to the original coordinates $( t, x ,y )$ given by
\eqn\rela{
u={ t\over y }\ ,\ \ \ v={ t\over
y }( x ^2+y ^2)\ ,\ \ \ \ X={ t x \over  y}\ .
}
It should be noted that it is not sufficient to add this
orbifold identification \pari\ to the string model of \LMS\ in
order to  have the fundamental domain $\F $ as constant time
slices. In addition, one has the restriction $uv-X^2\geq 0$. So
one can distinguish two related but different models: the
``null orbifold with reflection"  or just ``$SL(2,{\bf Z})$ orbifold",
obtained simply by adding
the reflection identification
\pari\ to the null orbifold model of \refs{\HS,\LMS }, and the ``restricted $SL(2,{\bf Z})$ orbifold model" \metrica\
having the finite volume
space  $\F $ as constant time slices. Both models arise as a quotient by the
$SL(2,Z)$ subgroup of the Poincar\'e isometry group
$SL(2,{\bf R})$ and have  $SL(2,{\bf Z})$ symmetry.
The former has another patch  $uv-X^2<0$ (i.e. $UV<0$), where the metric \abc\ can be put into the form
$ds^2=d t^2 + { t^2\ov
y ^2}(dx^2-dy^2)=  d t^2 + { t^2\ov
y ^2}  dz_+dz_- \ $, with $U=-t/y$, $V=ty$ and $z_\pm=x\pm y$.
In this patch, the identification \pari\ corresponds to the isometry
$z_+\to 1/z_+$, $z_-\to 1/z_-$, $t\to t$.
In the null orbifold model with reflection it seems more
convenient to work in $(u,v,X)$ coordinates. This model has
closed time-like curves arising in the region $uv-X^2<0$. To see
this, one can consider the image of  the point $u=v=0,\ X=1$
under an $SL(2,{\bf Z})$ transformation with parameters $\{
a,b,c,d\} $, $ad-bc=1$.\foot{We thank Hong Liu for providing this
example.}
{}For $bc>0$, the images $(u',v',X')=(2 ab, 2cd,1+2 bc)$ are time-like separated from the original point.
This transformation is a combination of reflection \pari\ and the
parabolic orbifold identification \idenf . The identification
\pari\ alone, or the identification  \idenf\ alone, do not produce CTC.

Here we shall use properties of modular functions in
$\F $ so in the remainder  we will consider the restricted $SL(2,{\bf Z})$ orbifold model. The two string models
should be  closely related.

In addition to the singularity at
$U= t/y =0$, the geometry  contains the singularities of $\F $ at $| z | =1, \
 x =\pm 1/2 $ for all $ t$.
 In the flat coordinates, it corresponds to
the subspace
$$
u=v=\pm 2X \ .
$$
There are no surviving supersymmetries,  they are broken by the
$SL(2,{\bf Z})$ identifications.

\smallskip

Let us now investigate  the string spectrum of this model and the structure of vertex operators. To leading order
in $\alpha '$, physical modes of the string spectrum obey, in a
suitable gauge, the Klein-Gordon equation. This applies to
massless and massive scalar fluctuations, as well as to non-zero
helicity components of a massive or massless mode. The only
exception are winding modes, which cannot be described in terms of
local fields of the low energy string effective action.
We will construct examples of invariant winding states at the end.

First, consider a massless scalar field propagating in this
geometry.
To exhibit the $SL(2,{\bf Z})$ invariance, it is
convenient to work in $( t, x ,y )$ coordinate
system. We  set the transverse
momentum components $p_i=0$.
The Laplace equation $\Delta_3 \Psi =0$ is
\eqn\lapl{
\Delta_2\Psi = \partial_t \big( t^2 \partial_t \Psi \big)\ ,
 }
$$
\Delta_2=y ^2( {\partial^2\over\partial x ^2}+
{\partial^2\over\partial y ^2})\ .
$$
We can solve it by separation of variables,
$\Psi =\psi_s( x ,y )\varphi_s( t)$, with
\eqn\jjj{
\Delta_2\psi_s=s(s-1)\psi _s\ .  \
}
This implies that the solution is given in terms of Maass waveforms.
These are modular functions which are eigenfunctions of the Laplace
 operator $\Delta_2 $ with
at most polynomial growth at $y \to\infty $  \terras .

Let us first  obtain the wave function by simply starting with the flat solution and summing over images.
In the coordinates $(u,v,X)$ the solutions of the
wave equation $\Delta_3 \Psi =0$ are plane waves with momentum
$(p_u,p_v,p_X)$, with $4p_up_v=p_X^2$.
Then a invariant wave function would have the generic form
\eqn\invas{\eqalign{
\Psi &=\sum_{\rm images} \int dp_udp_v dp_X
\ \varphi(p_u,p_v,p_X)\ e^{ip_vu+ip_uv+ip_X X}\cr
&=
\int dp_udp_v \  \tilde \varphi(p_u,p_v)\sum_{a,b,c,d}{}'\  \
e^{i{t\over y} p_u\big| (a+c \chi )z+b+d \chi \big| ^2}\ ,\ \ \ \chi=\sqrt{p_v/p_u}\ ,
\cr }
}
where $\sum '$ stands for a sum over integers $a,b,c,d$ obeying $ad-bc=1$, and we have used the relation \rela .
Normalizability of the wave function \invas\ (with the invariant measure $dxdy/y^2$) is a delicate issue.
To illustrate this point,
consider plane waves with $(p_u,p_v)=q(m^2,n^2)$, with $p_X$ again fixed by the mass
shell condition $4p_up_v=p_X^2$. They are of the form
\eqn\esf{
\Psi=
\int dq \ c(q) \ \sum_{(m,n)\neq (0,0)} e^{iq{ t\over y }|m+n z|^2}
\ .
}
Now consider a $c(q)$ of the form: $c(q)=c_s\ q^{s-1} \theta(q)$, with $s$ being
a complex number and $\theta(q)$ is the step function.
The
integral in $q$ is a Mellin transform. Computing this integral, one obtains
\eqn\invae{
\Psi = 2c_s \Gamma(s)\zeta(2s)\   t^{-s}\ E_{s}( z )\ , }
where $E_s$ is  the non-holomorphic
Eisenstein series
\eqn\eis{
E_s={1\over 2\zeta(2s)} \
\sum_{(m,n)\neq (0,0)}{y ^s\over |m+n z|^{2s}}\ .
}
It obeys the functional relation $E_s=\const\  E_{1-s}$,
so with no loss of generality one can restrict to ${\rm Re}\ s\geq 1/2$.\foot{ $E_s$ has a pole at $s=1$.
 $E_{s}$ with $s=1/2$  has a zero which is cancelled in \invae\ by the factor $\zeta(1)$, leading to  a $y^{1/2}\log y$ behavior at $y=\infty $.}
At $y=\infty $, $E_s$ behaves as $y^s$.
Therefore none of these wave functions \invae\ are   normalizable.

Normalizable wave functions can be written in terms of cusp forms,
which are automorphic functions that decrease exponentially at infinity.
They are bounded in $H_2$.
Let us summarize some known facts, referring to \terras\ for
details:
\smallskip

\noindent i) If ${\rm Re}\ s>1/2$ and $s\neq 1$, then the vector space ${\cal V}_s$ of Maass waveforms with eigenvalue $s (s-1)$
is generated by the non-holomorphic Eisenstein series $E_s$.  For $s=1$, the vector space ${\cal V}_s$ is generated by the constant.

\smallskip

\noindent ii) If  ${\rm Re}\ s=1/2$, the vector space ${\cal V}_s$ is generated
by $E_s$ and the cusp forms $v_n$, $n\geq 1$.
\smallskip

\noindent iii) The vector space of cusp forms $v_n$
is different from $\{ 0 \} $  only for eigenvalues  $s(s-1)$  with ${\rm Re}\ s=1/2$, for an infinite
number of values $s$. The set of values is discrete.
However, an explicit basis is not known, nor an explicit analytic example of a cusp form
(they can be easily found numerically).
\smallskip
Thus  cusp forms constitute the discrete part of the spectrum, while
$E_s$ constitute a continuous part of the spectrum.

A general eigenfunction in $L^2(SL(2,{\bf Z})\setminus H)$ is given in terms of the
Roelcke-Selberg expansion, which is the analog of a Fourier expansion in $\F $. Define the inner product
in the standard way as $(f,g)=\int_{\F }  \bar g(\bar z) f(z) y^{-2} dx dy$.
Then any $\psi$ in $L^2(SL(2,{\bf Z})\setminus H)$ has the expansion
\eqn\selb{
\psi (z)=\sum_{n\geq 0} (\psi ,v_n) v_n  + {1\ov 4\pi i} \int_{{\rm Re}\ s=\ha } (\psi ,E_s) E_s(z) ds\ ,
}
where $v_0=\sqrt{3/\pi}$.

Any cusp form is part of a vertex operator of the string model
having  non-zero quantum numbers in $\F $. A similar discussion
applies for the wave function of a massive scalar field: we solve
the equation \eqn\mav{ (\Delta_3 -M^2)\Psi =0 \ , } by separation
of variables, $\Psi= \psi_s( x ,y )\phi _s( t)$. The function
$\psi( x ,y )$ is then an eigenfunction of the Laplace operator
on $\F$, so it is also given in terms of cusp forms. Using eq.
\jjj , $\phi_s$ is then determined by \eqn\zzz{ \big[ \del_t t^2
\del_t - s(s-1)+M^2 t^{2}\big] \phi_s(t) = 0\  . } Hence
$\phi_s(t)=J_{\pm (s-\ha )}(Mt)/\sqrt{t} $.

 The cusp form can be expanded in terms of Bessel functions as follows:
\eqn\wop{
\psi_s(z)=\sum_{n\neq 0} a_n \ y^{1/2} K_{s-\ha }(2\pi |n|y)\exp(2\pi i n x)\ ,\ \ \ \ s=\ha+i b\ ,\ \ b\in{\bf R}\ ,
}
where $a_n$ are constrained by eq.~\jjj . Alternatively, one can make calculations in the covering space and
compute S-matrix elements by using plane-wave vertex operators of flat space and adding the images.
However,  we have seen above that naive vertices constructed by a sum over images as in \invas \ may not be normalizable (e.g.
 $c(q)=\sum_{k=0}^\infty c_k q^k $ does not give a normalizable wave function  \esf ).
$SL(2,{\bf Z})$ invariance severly constrains the form of vertices and amplitudes. In what follows we shall deduce a few interesting facts about the spectrum using theorems about cusp forms.

\smallskip

The ten dimensional string theory based on the $SL(2,{\bf Z})$ orbifold
has additional free directions $(x_1,...,x_7)$.
For a massless scalar fluctuation obeying $\Delta_{10}\Psi=0$,
with no string oscillations in the $t,x,y$ directions, the vertex is of the form
\eqn\ppss{
\Psi=\int d^7p\   a(p)  \ \phi_s(t) \ \psi_s (x,y)  \ \del x . \bar\del x\ e^{ip_i x_i}\ \ ,
}
where $\phi_s(t)$ is determined by \zzz\ with $M^2\to M^2+p_i^2$ (in this case, $M^2=0$).
In the frame $p_i=0$, the equation is of the form $\del_t t^2\del_t \phi_e=e \phi_e$.
The parameter $e$, being the eigenvalue of the time-derivative part of the Laplace operator,
measures the energy scale of the Kaluza-Klein excitation (bearing in mind that
in general  energy is not conserved since the background is not static).\foot{For
massless excitations, the energy parameter $e$ is associated with the scaling $t\to \Lambda t$.}
The spectrum formula is just given by
\eqn\espe{
 e =s (s-1 )\ ,\ \
} and $\phi_s(t)=a_s t^{-s}+b_s t^{s-1}$. {} For $s=0$ we have
$e=0$, corresponding to a state with trivial quantum numbers in
$\F $, being the only normalizable state with ${\rm Re}\ s\neq
1/2$. {}For other states, normalizability requires that $s=1/2 +i
b$, so that $e=-1/4-b^2$ is negative definite. Which is the value
of $e$ for the first non-trivial excitation? A theorem \terras \
establishes that $s (s-1)\leq -3\pi ^2/2$. This implies that the
energy $e$ of the first Kaluza-Klein state obeys the bound $e\leq
-3\pi ^2/2$. Precise numerical estimates of the first and the next
eigenvalues are summarized in \terras . Because the spectrum
contains both a discrete as well as a continuum part, one should
not expect that at low energy the theory will look $D-2$
dimensional.

Another important theorem establishes that the vector space of cusp forms with a given eigenvalue is
finite dimensional. This means that there is a finite number of normalizable excitations at each energy level $e$.

Finally, property iii) implies that the levels for cusp forms are discrete and that the
total number of normalizable excitations is infinite.
There is also an asymptotic formula from which one can estimate the total number
$N(e_0)$ of normalizable excitations
above an energy level $e_0$ (or with $|e|<|e_0|$),
 counted with multiplicity. For large $|e_0|$, one finds the remarkable formula
$N(e_0)\cong |e_0|/12$. This is a consequence of the Selberg trace formula \terras .

Similar results apply to quantum numbers of massive string modes.

\medskip

Let us now construct physical winding states associated with the twisted sector of the orbifold identifications \idenf , \pari .
In the coordinates $(u,v,X)$, the solution to the string equations
are expressed in terms of free right and left moving oscillators.
We only need to impose boundary conditions and solve the constraints.
We first consider the winding solution ($\s_\pm =\s\pm\tau $)
\eqn\winn{
u=p_v\tau\ ,\ \ \ \ \ X={np_v\over 8\pi }(\s_+^2-\s_-^2)={np_v\over 2\pi }\s \tau\ ,\ \ \ \ X(\s+2\pi )=X(\s )+n\ u\ .
}
Then  $v=v_+(\s_+)+v_-(\s_-)$ is determined by solving the constraint equations $T_{++}=T_{--}=0$. This gives
\eqn\vvv{
v= {n^2p_v\over 24\pi^2}\big( \s_+^3-\s_-^3 \big)={n^2p_v\over 12\pi^2}\big(\tau^3+3\s^2\tau \big)\
\ .
 }
$$
v(\s+2\pi )=v(\s )+n^2 u(\s)+2nX(\s )\ .
$$

In terms of $(t,x,y)$ coordinates, the solution takes the form
\eqn\txy{
t= {np_v\over 2\sqrt{3}\pi}\ \tau^2  \ , \ \ \ \ x={n\s \over 2\pi }\ ,\ \ \  y={n \over 2\sqrt{3}\pi}\ \tau\ .
}
{}For this state to be $SL(2,{\bf Z})$ invariant in $H_2$, one adds all the images in the upper half plane.

Next, consider the following solution which is a winding state for the twisted sector of \pari :
\eqn\abra{
X=2L \sin(\s/2)\cos(\tau/2)\ ,\ \ \ X_1= 2L\cos(\s/2)\cos(\tau/2)
\ ,\ }
$$
  u=X_0-X_1\ ,\ \ \ \ v=X_0+X_1\ .
$$
{}From the constraints, one gets $X_0=L\tau $. This solution
satisfies the boundary condition
$$
u(\s+2\pi )=v(\s )\ ,\ \ \ \ v(\s+2\pi )=u(\s )\ ,\ \  \ \ X(\s+2\pi )=-X(\s )\ .
$$
In the covering space, it describes a pulsating circular string loop, which contract to zero at $\tau=\pi $ and has maximum radius at $\tau=0, 2\pi $, etc.

In the  $(t,x,y)$ coordinate system, the solution becomes
$$
t=L\sqrt{\tau^2-4\cos^2(\tau/2) }\ ,\ \ \
$$
\eqn\bec{
x={2\sin(\s/2)\cos(\tau/2)\over \tau-2 \cos(\s/2)\cos(\tau/2)}\ ,\ \ \ y={\sqrt{\tau^2-4\cos^2(\tau/2) }\over
\tau-2 \cos(\s/2)\cos(\tau/2)}\ ,
}
with the addition of the images.
Note that in these coordinates the solution exists after a $\tau_0\cong 1.48$
such that $\tau^2-4\cos^2(\tau/2)>0$. For $\tau<\tau _0$, the condition $uv-X^2>0$ is not satisfied for all points of the string, i.e. part of the string is in the patch $uv-X^2<0$
 which is not covered by the $(t,x,y)$ coordinates.
Different points of the string pass through the singularity $u=0$ at different times $X_0=L\tau $.
 As $\tau $ gradually increases
and goes over $\tau_0$, the string fully enters into the region $uv-X^2>0$.
At this point, the string has already  left the singularity $u=0$.
 In the upper half plane,
when $\tau $
is slightly above $\tau_0$, the string describes a long loop.
When $\tau=\pi $, it contracts to a point at $x=0, y=1$. Then it keeps pulsating around this point,
with a maximum radius that decreases with time.

In the Hilbert space, these  classical string configurations \txy ,\bec\ are described by physical
string states with large quantum numbers, so that the classical description applies.
This requires large $p_v $ in the case of \winn , and large $L$ --~which implies large occupation number of the
lowest frequency string oscillator~-- in the case of \abra , \bec .
The state \txy\ which winds around $ x $ becomes ``massless" near
$U=0$ (the light-cone energy $p_u$ is proportional to $U^2n^2/p_v$, see \abc , \nuev ). This is a reflection of the
singularity of the geometry at $U=0$.

The instabilities pointed out in \HP\ may be absent in the $SL(2,{\bf Z})$ orbifold model.
Indeed, this instability originated from
the gravitational interaction of plane waves and their images.
Although here the spectrum has also a continuum part which may lead to wave interactions,
this part is severely
restricted by $SL(2,Z)$ symmetry, so
 the argument of \HP\
does not seem to directly apply to this model. {}The  states in
the discrete part of the spectrum have finite motion. The
corresponding wave functions are regular (bounded) on $\F $,
though they exhibit a singular behavior at $t=0$. It would be
interesting to investigate  the partition function and higher
point scattering amplitudes to establish if the string model is
regular or singular.
 In the $SL(2,Z)$ orbifold model,  the extra reflection symmetry \pari \ adds
 new terms to the partition
function computed in \LMS .

\bigskip

\noindent {\bf Acknowledgements}

\noindent We thank Joaquim Gomis and Hong Liu  for
useful discussions and comments.
This  work is
supported in part by the European Community's Human Potential Programme
under contract HPRN-CT-2000-00131, and by MCYT
FPA, 2001-3598 and CIRIT GC 2001SGR-00065.

\vfill\eject \listrefs
\end